\documentclass[structabstract]{aa}  

\usepackage[fleqn]{amsmath}
\usepackage{amssymb}
\usepackage[varg]{txfonts}
\usepackage{graphicx}
\usepackage{natbib}
\usepackage{booktabs}
\usepackage{array}
\usepackage{fancyhdr}
\usepackage{multirow}

\bibpunct{(}{)}{;}{a}{}{,}
\def\apjl{ApJ}%
\def\apjs{ApJS}%
\def\ao{Appl.~Opt.}%
\def\aap{A\&A}%
\renewcommand{\vec}[1]{\ensuremath{\mathchoice{\mbox{\boldmath$\displaystyle#1$}}
{\mbox{\boldmath$\textstyle#1$}}
{\mbox{\boldmath$\scriptstyle#1$}}
{\mbox{\boldmath$\scriptscriptstyle#1$}}}}
\begin{document}
   \title{Maximum likelihood estimation of local stellar kinematics}

   \author{T. Aghajani
          \and
          L. Lindegren
          }

   \institute{Lund Observatory, Lund University, Box 43, 22100 Lund, Sweden\\
              \email{toktam.aghajani@gmail.com, Lennart.Lindegren@astro.lu.se}
             }

\date{Received September 21, 2012 / Accepted November 28, 2012}

 \abstract
 % context heading (optional), leave it empty if necessary  
  {Kinematical data such as the mean velocities and velocity dispersions of stellar samples
are useful tools to study galactic structure and evolution. However, observational data are often
incomplete (e.g., lacking the radial component of the motion) and may have significant 
observational errors. For example, the majority of faint stars observed with \emph{Gaia} will not
have their radial velocities measured.}
 % aims heading (mandatory)
  {Our aim is to formulate and test a new maximum likelihood approach to estimating the kinematical 
parameters for a local stellar sample when only the transverse velocities are known (from parallaxes 
and proper motions).
}
 % methods heading (mandatory)
  {Numerical simulations using synthetically generated data as well as real data (based on the
Geneva--Copenhagen survey) are used to investigate the statistical
properties (bias, precision) of the method, and to compare its
performance with the much simpler ``projection method'' described by Dehnen \& Binney (1998).
}
 % results heading (mandatory)
  {The maximum likelihood method gives more correct estimates of the dispersion when observational 
errors are important, and guarantees a positive-definite dispersion matrix, which is not always
obtained with the projection method. Possible extensions and improvements of the method are discussed.
}
 % conclusions heading (optional), leave it empty if necessary 
   {}


   \maketitle
%
%________________________________________________________________

\section{Introduction}

Statistical information about the motions of stars relative to the sun
may contain important hints concerning the origin and history of the
stars themselves (e.g., by identifying kinematic populations and
streams) as well as of the physical properties of the Galaxy (through
dynamical interpretation of the motions). Ideally this requires that
all six components of phase space (positions and velocities) are known
for all the stars in the investigated sample. This can in principle be
achieved through a combination of astrometric data (providing
positions and distances, from the parallaxes, and tangential
velocities from the proper motions and parallaxes) and spectroscopic
radial velocities. Very often, however, such complete data are not
available for all the stars in a sample. For example, the
\emph{Hipparcos}
Catalogue \citep{hip:catalogue} gives the required astrometric
information for large samples of nearby stars, but not all of them
have known radial velocities. Restricting the investigation to stars
with measured radial velocities could introduce a serious selection
bias \citep{1997ESASP.402..473B}. The \emph{Gaia} mission
\citep{2012Ap&SS.tmp...68D} will not only provide vastly improved
astrometric data for much larger stellar samples, but also
radial-velocity measurements for stars brighter than $\simeq 17$~mag;
for the fainter stars, however, the phase space data will still be
incomplete. On the other hand, on-going spectroscopic surveys such as
RAVE \citep{2006AJ....132.1645S} already provide radial velocities for
large samples without the complementary astrometry. Thus we are often
faced with the problem to estimate kinematical parameters (such as
mean velocities and velocity dispersions) from incomplete phase space
data, lacking either the radial or tangential velocity components, or
the accurate distances needed to derive the tangential velocities from
proper motions.

\citet{DB98} derive a simple and elegant method to
derive local stellar kinematics (mean velocity and velocity
dispersion) for a group of stars, when the tangential velocities are
known, but not the radial velocities. This method has become popular
and can be considered a standard for this purpose. We
will refer to it as the \emph{projection method} (PM) throughout this paper.
In addition to the original work by \citet{DB98}, the same method (or variants of it) has
been used, e.g., by \citet{Mignard2000}, \citet{Brosche+2001},
\citet{F.Leeuwen2007}, and \citet{Aumer+Binney2009}. 

Despite its wide usage in the literature, the projection method is not
founded on any particular estimation principle such as maximum
likelihood (ML) or Bayesian estimation, but simply fit the projected
first and second moments of the space velocities to the corresponding
observed moments of the tangential velocities. This works well for
large samples, provided that the observational uncertainties in the
data are negligible compared to the uncertainties due to the sampling,
but there is no guarantee that it is unbiased, or asymptotically
efficient as expected for an ML estimate. On the contrary, by
neglecting the observational errors the resulting velocity dispersions
are probably overestimated. Moreover, for small samples the
projection method may sometimes give unphysical results, in that the
estimated dispersion tensor is not positive-definite -- implying that
the mean square velocity is negative in some directions. Since it is
desirable that the kinematic parameters can be consistently and
efficiently estimated also for small samples and in the presence of
non-negligible observational uncertainties, we introduce here a new
and more rigorous approach, based on the maximum likelihood method.

%________________________________________________________________

\section{Kinematic parameters and the projection method}

For a homogeneous population of stars the phase space density
$f(\vec{r},\vec{v},t)$ describes the density of stars as a function of
position ($\vec{r}$), velocity ($\vec{v}$) and time $t$. By the
\emph{local kinematics} we mean the distribution function here
($\vec{r} = \vec{0}$) and now ($t = 0$), that is $f(\vec{v}) \equiv
f(\vec{0},\vec{v},0)$. It is usually assumed that  $f(\vec{v})$ is a
smooth function, and the most common assumption is the Schwarzschild
approximation, that $f(\vec{v})$ is a three-dimensional Gaussian
distribution (velocity ellipsoid), or a combination of a few Gaussian
distributions. The velocity ellipsoid is completely described by the
mean velocity $\vec{v}$ and the dispersion tensor $\vec{D}$.

Throughout this paper we use heliocentric galactic coordinates $(x, y,
z)$ with $+x$ pointing towards the Galactic Centre, $+y$ in the
direction of rotation at the Sun, and $+z$ towards the north Galactic
pole. The corresponding heliocentric velocity components are denoted
$(v_x,v_y,v_z)$ or $(u,v,w)$.

For a stellar population, the mean velocity and dispersion tensor are defined as
\begin{equation}
\overline{\vec{v}} = \text{E}\left[ \vec{v} \right] \, ,
\label{eq:v}
\end{equation}
\begin{equation}
\vec{D} = \text{E}\left[ (\vec{v}-\overline{\vec{v}})
  (\vec{v}-\overline{\vec{v}})^\text{T} \right] \, ,
\label{eq:D}
\end{equation}
where E is the expectation operator (population mean). Given the
3-dimensional velocities $\vec{v}_i$, $i=1,~2,~\dots,~n$ for a sample
of stars it is possible to estimate the population mean values by
the sample mean values,
\begin{equation}\label{e00}
\overline{\vec{v}} \simeq \frac{1}{n}\sum_i \vec{v}_i\, , \quad
\vec{D} \simeq \frac{1}{n}\sum_i (\vec{v}_i-\overline{\vec{v}})(\vec{v}_i-\overline{\vec{v}})^\text{T}\, .
\end{equation}
Note that this estimate of $\vec{D}$ is always positive definite
(except in trivial degenerate cases), but it requires 3-dimensional
velocities and the dispersions are likely to be overestimated when the
velocity components have significant observational errors.

The tangential velocity of a star can be written
\begin{equation}\label{eq:tau}
\vec{\tau} = \vec{A}\vec{v} \, ,
\end{equation}
where $\vec{A}$ is a projection matrix depending only on the position
of the star. In the projection method the assumption that the
positions (and hence the matrices $\vec{A}$) are uncorrelated with the
velocities is invoked to derive a relation between the mean of
$\vec{\tau}$ and the mean of $\vec{v}$, allowing the latter to be
solved. In a similar way, the elements of $\vec{D}$ are derived from
the relation between the second moments of $\vec{\tau}$ and
$\vec{v}$. For further details we refer to the paper by Dehnen \&
Binney (1998).

%________________________________________________________________

\section{Maximum likelihood estimation of the kinematic parameters}

\subsection{Overview of the model} \label{sec:model}
We assume a Gaussian distribution of the velocities
in all directions. That is, the
velocity of a star follows a 3-dimensional normal distribution,
$\vec{v} \sim  \vec{N}(\overline
{\vec{v}},\vec{D})$.
\footnote{This notation $\vec{x}\sim\vec{N}(\vec{m},\vec{V})$
means that the random variable $\vec{x}$ follows the multi-dimensional
normal distribution with mean value $\vec{m}$ and covariance matrix 
$\vec{V}$. Similarly for a one-dimensional normal variable 
$x\sim N(m,s^2)$, $m$ is the mean value and $s^2$ the variance.} 
Given the position and true parallax $p$ of the star we can calculate its 
true proper motion in longitude and latitude
$(\mu_{l},\mu_{b})$. Adding Gaussian observational errors to these we get the observed proper
motions $\tilde{\mu_{l}} \sim  N(\mu_l,\sigma_{\mu}^2)$,
$\tilde{\mu_{b}} \sim  N(\mu_b,\sigma_{\mu}^2)$ and the observed
parallax $\tilde{p} \sim  N(p,\sigma_p^2)$. The observational
uncertainties $\sigma_\mu$ and $\sigma_p$ are assumed to be known. The
ML formulation requires that the probability density function of the
observed data $(\tilde{\mu}_l,\tilde{\mu}_b,\tilde{p})$ is written as a
function of the model parameters.

\subsection{Exact expression for the likelihood}
For a problem involving $n$ stars, the parameters of the model are:
\begin{itemize}
\item
$\overline{\vec{v}}=$ the mean velocity of the stellar population (a
3-vector, or $3\times 1$ matrix);
\item $\vec{D}=$ the dispersion
tensor of the stellar population (a symmetric $3\times 3$ matrix; contains 6 non-redundant elements);
\item $\vec{p}=$ the true parallaxes of the stars (an $n$-vector, or $n\times 1$ matrix).
\end{itemize}
It is necessary to introduce the true parallaxes $p_i$ as formal model
parameters, although the strategy is that they will be eliminated on a
star-by-star basis leaving us with a problem with only nine model
parameters, namely the (non-redundant) components of $\overline{\vec{v}}$
and $\vec{D}$. We denote by the vector $\vec{\theta}$ the complete set
of model parameters (i.e., the unknowns to be estimated).
For $n$ stars the total number of model parameters is $n+9$. 

The observables are, for each star $i=1\dots n$, the observed components
of proper motion, $\tilde{\mu}_{l,i}$ and $\tilde{\mu}_{b,i}$, and the
observed parallax $\tilde{p}_i$. The total number of observables is
$3n$. These have observational uncertainties that are given by the
$3\times 3$ covariance matrix $\vec{C}_i$. Sometimes it is useful to
denote by the vector $\vec{x}$ the complete set of observables (or
data).

Given the observations, the likelihood function $L(\vec{\theta})\equiv
L(\overline{\vec{v}},\vec{D},\vec{p})$ numerically equals the probability
density function (pdf) $f_{\vec{x}}(\vec{x}|\vec{\theta})$ of the
observables $\vec{x}$, given the model parameters
$\vec{\theta}$. The objective is to find the ML estimate of $\vec{\theta}$, denoted $\hat{\vec{\theta}}$, i.e., the
(hopefully unique) set of parameter values that maximizes
$L(\vec{\theta})$ or, equivalently, the log-likelihood
$\ell(\vec{\theta}) = \ln L(\vec{\theta})$.
The total log-likelihood function is given by
\begin{equation}\label{e01}
\ell(\overline{\vec{v}},\vec{D},\vec{p}) = \sum_i \left[ \ln f_{\tilde{\vec{\mu}},i}(\tilde{\vec{\mu}}_i|p_i)
+ \ln g_i(\tilde{p}_i-p_i) \right] \, ,
\end{equation}
where $g_i$ is the centered normal pdf with standard deviation $\sigma_{p,i}=[\vec{C}_i]_{33}^{1/2}$,
i.e., the uncertainty of the parallax $p_i$. It is clear that the
parameter $p_i$ only affects the $i$th term in the sum
above. Therefore, when maximizing with respect to $p_i$ we only need
to consider that one term. For simplicity we drop, for the moment,
the subscript $i$ so that the term to consider (for one star) can be
written:
\begin{align}\label{e02}
\ell(\overline{\vec{v}},\vec{D},p) &= \ln f_{\tilde{\vec{\mu}}}(\tilde{\vec{\mu}}|p) + \ln g(\tilde{p}-p)
\nonumber \\
&= \ln f_{\tilde{\vec{\mu}}}(\tilde{\vec{\mu}}|p)-\frac{1}{2}\ln(2\pi\sigma_p^2)
-\frac{(p-\tilde{p})^2}{2\sigma_p^2} \, ,
\end{align}
where we have used $g(x)=(2\pi\sigma_p^2)^{-1/2}\exp(-x^2/2\sigma_p^2)$ 
for the normal pdf with standard deviation $\sigma_p$. Note
that $f_{\tilde{\vec{\mu}}}(\tilde{\vec{\mu}}|p)$ of course depends on
$\overline{\vec{v}}$ and $\vec{D}$ as well, although we focus on the
dependence on $p$ here. We now need an explicit expression for the pdf
$f_{\tilde{\vec{\mu}}}(\tilde{\vec{\mu}}|p)$ of the observed proper
motion $\tilde{\vec{\mu}}$ when the true parallax $p$ is known. This is
clearly a two-dimensional Gaussian (since both the velocities and the 
observational errors are Gaussian), with expected value
\begin{equation}\label{e101}
\text{E}\left[ \tilde{\vec{\mu}} \right] = \vec{M}\overline{\vec{v}} \, .
\end{equation}
Here, $\vec{M}$ is the $2\times 3$ matrix
\begin{equation}\label{e102}
\vec{M} = \frac{p}{K}
\begin{bmatrix} -\sin l & \cos l & 0 \\
-\cos l\sin b & -\sin l\cos b & \cos b 
\end{bmatrix}
\end{equation}
projecting any vector (in this case $\overline{\vec{v}}$) to its components
in the directions of increasing galactic coordinates $l$ and $b$.
The factor $p/K$ converts the velocity to proper motion (numerically,
$K=4.7405$ if the units used are km~s$^{-1}$, mas, and mas~yr$^{-1}$).
Assuming that the observational errors are not correlated with the 
velocities, the covariance of $\tilde{\vec{\mu}}$ is the sum of the covariance 
due to the velocity dispersion and the covariance due to the observational 
errors, i.e.:
\begin{equation}\label{e103}
\text{Cov}\left[ \tilde{\vec{\mu}} \right] \equiv \vec{C}_{\tilde{\vec{\mu}}}
= \vec{M}\vec{D}\vec{M}^{\rm T} + \begin{bmatrix} \sigma_{\mu l}^2 &
\rho\sigma_{\mu l}\sigma_{\mu b} \\
\rho\sigma_{\mu l}\sigma_{\mu b} & \sigma_{\mu b}^2 \end{bmatrix},
\end{equation}
where $\sigma_{\mu l}$ and $\sigma_{\mu b}$ are the observational
uncertainties in $\mu_l$ and $\mu_b$, and $\rho$ their correlation
coefficient. The pdf is then
\begin{equation}\label{e104}
f_{\tilde{\vec{\mu}}}(\tilde{\vec{\mu}}|p) = (2\pi)^{-1} 
\left| \vec{C}_{\tilde{\vec{\mu}}} \right|^{-1/2} 
\exp\left[ -\frac{1}{2}\left(\tilde{\vec{\mu}}-\vec{M}\overline{\vec{v}}\right)'
\vec{C}_{\tilde{\vec{\mu}}}^{-1}\left(\tilde{\vec{\mu}}-\vec{M}\overline{\vec{v}}\right)
\right] \, .
\end{equation}

%________________________________________________________________

\subsection{Eliminating the parallaxes}

The large number of parameters in the likelihood function
in Eq.~(\ref{e01}) makes it difficult to maximize. We approach
this problem by finding the maximum of
Eq.~(\ref{e02}) with respect to $p$ for each star and thus
eliminating $\vec{p}$. Since the expression for
$f_{\tilde{\vec{\mu}}}(\tilde{\vec{\mu}}|p)$ is quite complicated, it
is not likely that this can be done exactly, by analytical means, but
we can find an approximate solution valid in the limit of small
relative parallax error, $\sigma_p/p \ll 1$. 

To eliminate $p$ we take the partial derivative of Eq.~(\ref{e02}) with 
respect to $p$ and set it equal to 0. The result is 
\begin{equation}\label{e04}
p = \tilde{p} + \sigma_p^2 \, F(p)\, ,
\end{equation}
where we introduced for brevity
\begin{equation}\label{e06}
F(p) \equiv \frac{\partial\ln f_{\tilde{\vec{\mu}}}(\tilde{\vec{\mu}}|p)}{\partial p} \, .
\end{equation}
So far no approximation has been made. However, in Eq.~(\ref{e04}) 
the function $F$ is to be evaluated for the value of $p$ which we are
seeking, so we have not really solved the problem. But if
$\sigma_p \ll p$ we can evaluate $F(p)$ at $p = \tilde{p}$ 
to obtain the explicit formula
\begin{equation}\label{e07}
p  \simeq \tilde{p} + \sigma_p^2 \left ( \frac{\partial\ln
  f_{\tilde{\vec{\mu}}}(\tilde{\vec{\mu}}|p)}{\partial p} \right )_{p
= \tilde{p}} = \tilde{p} + \sigma_p^2 \, F(\tilde{p}) \, .
\end{equation}
This approximation is safe to do under the assumption
of small relative parallax error, since
Eq.~(\ref{e07}) is obviously correct to first order in $\sigma_p^2$.
Inserting the first-order Taylor expansion
\begin{equation}\label{e08}
\ln f_{\tilde{\vec{\mu}}}(\tilde{\vec{\mu}}|p) \simeq 
\ln f_{\tilde{\vec{\mu}}}(\tilde{\vec{\mu}}|\tilde{p}) + (p-\tilde{p})F(\tilde{p})
\end{equation}
in Eq.~(\ref{e02}) and using $p-\tilde{p}\simeq\sigma_p^2F(\tilde{p})$
from Eq.~(\ref{e07}), the log-likelihood maximized with respect to $p$ becomes:
\begin{equation}\label{e11}
\ell(\overline{\vec{v}},\vec{D}) \simeq \ln f_{\tilde{\vec{\mu}}}(\tilde{\vec{\mu}}|\tilde{p}) 
+ \frac{1}{2}\sigma_p^2\,F(\tilde{p})^2 \, .
\end{equation}
Here we have neglected the additive constant
$-\frac{1}{2}\ln(2\pi\sigma_p^2)$, which does not depend on the
model parameters and therefore is irrelevant for the ML problem.

Recalling that this is just the log-likelihood term for one star, the
total log-likelihood function, after maximization with respect to
$\vec{p}$, is therefore:
\begin{equation}\label{e12}
\ell(\overline{\vec{v}},\vec{D}) \simeq \sum_i \left[ 
\ln f_{\tilde{\vec{\mu}},i}(\tilde{\vec{\mu}}_i|\overline{\vec{v}},\vec{D},\tilde{p}_i) 
+ \frac{1}{2}\sigma_{p,i}^2\,F_i(\tilde{p}_i)^2 \right].
\end{equation}

%________________________________________________________________

\subsection{Numerical solution method}

We make use of a numerical method for maximizing the likelihood
function, rather than an analytical one, as the derivatives of
Eq.~(\ref{e12}) become very complicated. We have used the Nelder--Mead
simplex method \citep{lagarias1998cpn}, implemented in MATLAB
as the function {\tt fminsearch}, to minimize the negative of the 
log-likelihood. The simplex method is useful as it does not require 
us to know the derivatives. 

As soon as there are enough stars in the sample the minimization 
works fine but we have found two situations in which the algorithm
diverges. The first happens when some stars in the
sample have very small parallaxes or very large tangential velocities. 
The second situation may occur when the sample is very small.
What happens in both cases is that the estimated dispersion
tends towards zero in some direction, apparently because the
measurement errors are enough to explain the observed dispersion, 
resulting in a singular dispersion matrix. 

To compute a solution even in these cases, we introduce a regularization parameter 
$\alpha > 0$. The regularized log-likelihood is:
\begin{equation}
\label{e13}
\ell(\overline{\vec{v}},\vec{D}) \simeq \sum_i \left[ 
\ln f_{\tilde{\vec{\mu}},i}(\tilde{\vec{\mu}}_i|\overline{\vec{v}},\vec{D},\tilde{p}_i) 
+ \frac{1}{2}\sigma_{p,i}^2\,F_i(\tilde{p}_i)^2 \right]- 
\alpha \ln \frac{S_{\rm max}}{S_{\rm min}},
\end{equation}
where $S_{\rm max}$ and $S_{\rm min}$ are the extreme singular values
of the singular value decomposition
of $\vec{D}$. $S_{\rm max}/S_{\rm  min}$ is the square of the ratio 
of the longest and shortest axes of the velocity ellipsoid.
 If $\alpha$ is large the algorithm will tend to favour a small axis
 ratio, resulting in an isotropic velocity dispersion in the limit of large
$\alpha$. In order to use as little regularization as necessary we
try Eq.~(\ref{e13}) with $\alpha = 0$ and, if diverging, increase
$\alpha$ by steps of 0.5 until a converged estimate is obtained. 
The result is a dispersion matrix that is always positive-definite, but 
when $\alpha>0$ it may be more isotropic than it should be
(Table~\ref{alphatable}).

%                                                                    One column table
%________________________________________________________________
\begin{table}[t]
\caption{An example with 100 stars illustrating how the estimated velocity dispersion depends on 
the value of $\alpha$.}
\centering
\scalebox{1}{
\begin{tabular}{ccccccc}
\toprule
$\alpha$ & \multicolumn{1}{c}{$\overline{u}$} &
\multicolumn{1}{c}{$\overline{v}$} & \multicolumn{1}{c}{$\overline{w}$} &
\multicolumn{1}{c}{$\sigma_u$} & \multicolumn{1}{c}{$\sigma_v$} &
\multicolumn{1}{c}{$\sigma_w$} \\
\cmidrule{1-7}
True & $10.000 $ & $15.000 $ & $7.000$ &22.000  & 14.000 & 10.000 \\ 
\midrule
0.0 & $ 10.090 $ & $14.775$ & $7.083$ & 21.712  & 13.636 & 9.783  \\ 
 0.5& $ 10.033$ & $14.808$ & $7.090$ & 21.610 & 13.638 & 9.902 \\ 
1.0& $10.032$ & $14.806 $ & $7.091$ & 21.429  & 13.619  & 10.041 \\
1.5 & $10.031$ & $14.805$ & $7.093$ & 21.251 & 13.604 & 10.182  \\
2.0 & $10.029$ & $ 14.804 $ & $7.094$ & 21.076 & 13.591 & 10.324\\
10.0 & $10.015$ & $14.824$ & $7.143 $ & 18.587 & 13.845 & 12.819  \\
\bottomrule
 \end{tabular}}
\tablefoot{
In this case $\alpha=0$ would actually be used (no regularization). The solutions 
for $\alpha>0$ were only made to show that the velocity dispersion becomes more isotropic when 
regularization is used.
}
 \label{alphatable}
\end{table} 

The derivative in Eq.~(\ref{e06}) is calculated numerically, using the 
approximation
\begin{equation}
 \dfrac{\partial \ln f_{\tilde{\mu}}(\tilde{\mu}|p)}{\partial p}\simeq
 \dfrac{f_{\tilde{\mu}}(\tilde{\mu}|p+\sigma_p)-f_{\tilde{\mu}}(\tilde{\mu}|p-\sigma_p)}
 {2\sigma_p} \, .
\end{equation} 
Using a step size of $\pm\sigma_p$ is logical as we are looking for solutions 
typically within one standard deviation of the observed value. 

%________________________________________________________________

\section{Testing on synthetic and real samples}

We test our method (ML) together with the projection method (PM) as described by
\citet{DB98} on two types of samples. The first type is a
synthetic sample, where nearby ``stars'' with known mean velocity and
velocity dispersion have been generated. This sample allows us to
investigate the bias of the two methods. The second type is a sample of
real stars taken from the Geneva--Copenhagen survey of nearby F and G
dwarfs \citep{2004A&A...418..989N}. These stars have measured radial
velocities which allow us to estimate their mean velocity and
velocity dispersion directly from Eq.~(\ref{e00}). 
We use the resulting estimates as the ``true'' values when comparing 
the methods.

%________________________________________________________________

\subsection{Synthetic samples}
\label{sec:synth}

%________________________________________________________________

\subsubsection{Generating the data}

The generation of synthetic stars strictly follows the probabilistic
model outlined in Sect.~\ref{sec:model}. Observational
uncertainties were set to $\sigma_\mu=1$, 3, 10 and 30~mas~yr$^{-1}$
for the components of the proper motions, and to $\sigma_p=1$~mas for the parallaxes.
The stars have randomly generated positions $(x,y,z)$ (in pc) within a radius of 100~pc 
from the sun. Each star is also given a randomized velocity
using a normal distribution with
the mean parameter $\overline{\vec{v}} = (10,~15,~7)$~km~s$^{-1}$ and
dispersion matrix $\vec{D} = \rm diag(22^2,~14^2,~10^2)$~km$^2$~s$^{-2}$. From the
rectangular coordinates the galactic longitude $l$ and galactic
latitude $b$ are obtained with
\begin{equation}
    l = \mbox{atan}2(y, x) \, , \quad
    b = \mbox{atan}2(z, \sqrt{x^2+y^2}) \, .
\end{equation}
Then for any position in sky $(l,b)$  we define three orthogonal unit vectors
\begin{equation}
\vec{u} =
\left[ 
\begin{array}{c}
\vspace{2pt}
\cos b \cos l \\
\vspace{2pt}
\cos b \sin l\\
\sin b 
\end{array} 
\right]
\quad
\vec{l} =
\left[ 
\begin{array}{c}
\vspace{2pt}
-\sin l   \\
\vspace{2pt}
\cos l\\
0 
\end{array} 
\right]
\quad
\vec{b} =
\left[ 
\begin{array}{c}
\vspace{2pt}
-\sin b \cos l   \\
\vspace{2pt}
-\sin b \sin l\\
\cos b
\end{array} 
\right],
\end{equation}
where $\vec{u}$ is the direction towards the star, and $\vec{l}$ and $\vec{b}$ 
are unit vectors in the tangential plane in the sky: $+\vec{l}$ in the direction of 
increasing longitude, and $+\vec{b}$ in the direction of increasing latitude. 
$[\vec{u},\vec{l},\vec{b}]$ is called the normal galactic triad at the point $(l,b)$. 

The true parallax is given by
\begin{equation}
p = \frac{1000~\textrm{mas}}{\sqrt{x^2+y^2+z^2}} \, ,
\end{equation}
whereupon the true proper motion in longitude and latitude is calculated as
\begin{equation}
\mu_l = \frac{p}{K}\,\vec{l}\cdot \vec{v} \quad\textrm{and} \quad \mu_b = \frac{p}{K}\,\vec{b}\cdot \vec{v}
\end{equation}
respectively. Here, $K = 4.7405$~km~s$^{-1}$~kpc$^{-1}$~(mas~yr$^{-1}$)$^{-1}$ is
the numerical constant relating mas~yr$^{-1}$ to
km~s$^{-1}$~kpc$^{-1}$.

To get the observed proper motions and parallax, the errors in proper
motion and parallax are added to the true values. Errors were
drawn from the normal distribution with zero mean value and standard
deviation equal to the observational uncertainties $\sigma_{\mu}$ and
$\sigma_{p}$.

As the simulated data set depends on many different parameters it is
not possible to investigate the methods for all possible
combination of the parameters. We restrict ourselves to keep most of the
parameters fixed ($\overline{\vec{v}}$, $\vec{D}$, $R$ and $\sigma_p$), and
only study how the methods perform as a function of $n$, the number of
stars in the sample, and $\sigma_\mu$, the uncertainty in proper
motion. We use $n=30$, 100, 300, 1000 and
$\sigma_\mu=1$, 3, 10, 30~mas~yr$^{-1}$, which gives 16 different
combinations. For each combination 100 samples are generated with 
different seeds for the random generator. The 100 samples are thus
different in the generated true positions and velocities, as well as in 
the observational errors. The PM and ML methods were applied
to all 1600 samples and the mean values and RMS variations of the
estimated parameters calculated.

%________________________________________________________________

\subsubsection{Results}
\label{sec:resultssynth}

With the synthetic sample we can see a difference in how the uncertainty in proper
motion affects the values of the calculated velocities and velocity
dispersions for the two methods. 

\begin{figure*}[tbp]
\begin{center}
\centerline{
\includegraphics[width= 0.82\columnwidth]{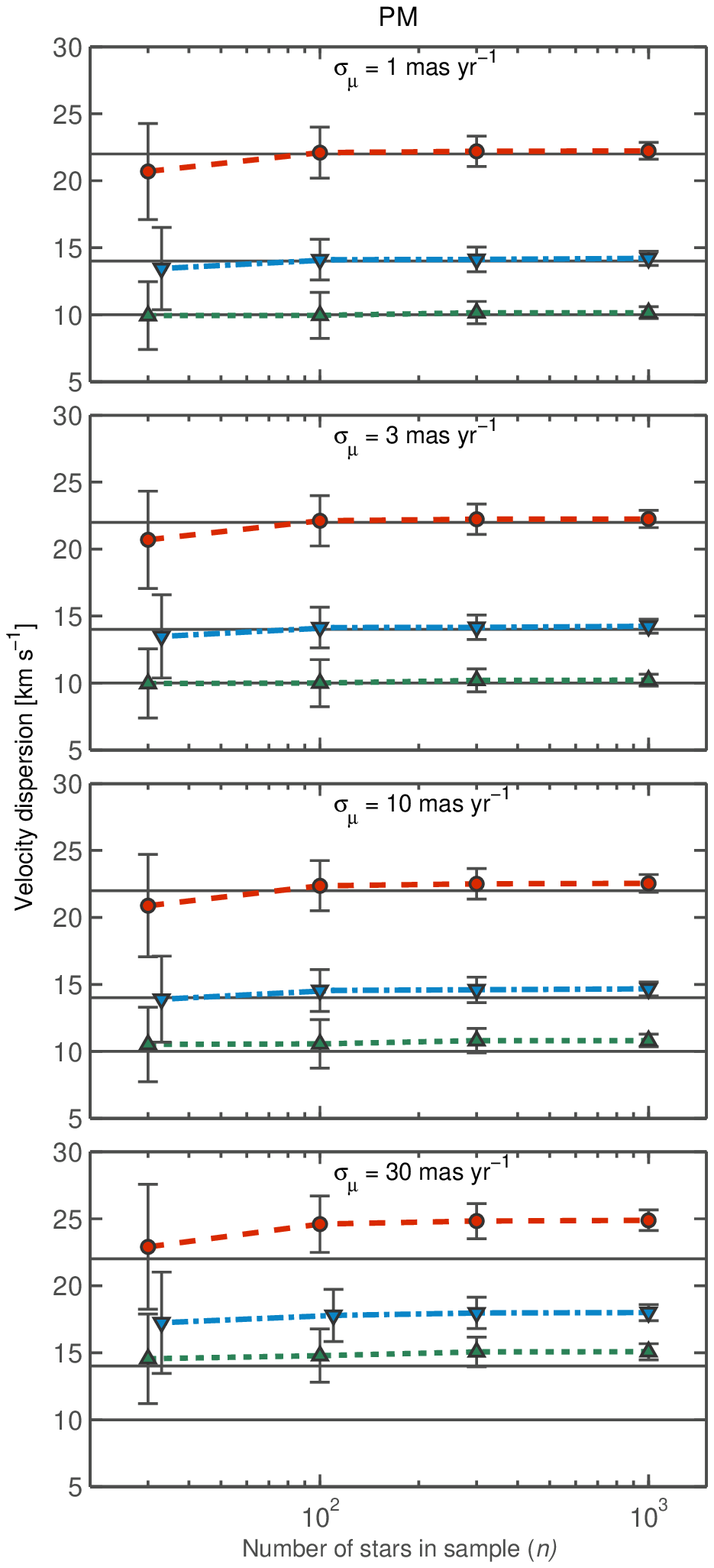}
\hspace{0.5cm}
\includegraphics[width= 0.82\columnwidth]{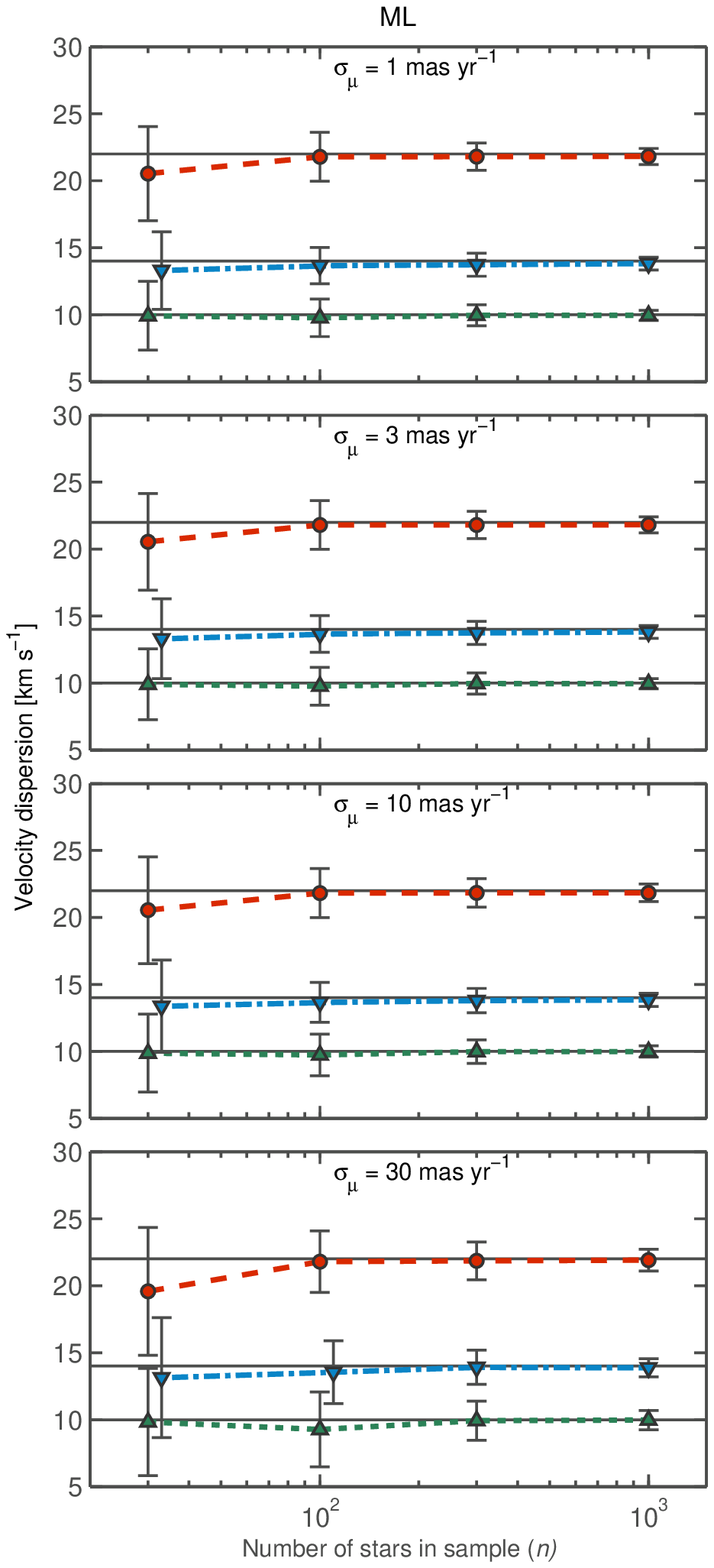}
}
\end{center}
\vspace{-.8cm}\caption{Plots illustrating velocity dispersion versus the number of stars, for different values of $\sigma_{\mu}$,
as calculated for the synthetic samples described in Sect.~\ref{sec:synth}.
Diagrams in the left column are for the projection method (PM), the right column is for the maximum likelihood
method (ML). The symbols represent the different galactic components of the dispersion:
$\sigma_u$ ($\bullet$), $\sigma_v$ ($\blacktriangledown$), $\sigma_w$ ($\blacktriangle$). 
The true values are indicated by the horizontal lines.}\label{fig:dispersionvsuncertainty}
\end{figure*}

In the left column of
Fig.~\ref{fig:dispersionvsuncertainty}, results from the PM are
shown. 
As expected, the dispersions are increasingly biased towards 
higher values when the uncertainty in proper motion increases. The effect is negligible
when $\sigma_\mu=1$ or 3~mas~yr$^{-1}$, but for $\sigma_\mu=10$ or
30~mas~yr$^{-1}$ the increase in the dispersions is quite
significant. 
This is reasonable since an error of 30~mas~yr$^{-1}$, at the median 
distance of the sample (80~pc), corresponds to a linear velocity error of about 11~km~s$^{-1}$, 
which is comparable to the true velocity dispersion. As can be seen in the bottom 
left panel of Fig.~\ref{fig:dispersionvsuncertainty}, the dispersion component in $w$ is most
severely overestimated ($\simeq\! 15$~km~s$^{-1}$ versus the true value 10~km~s$^{-1}$),
while the $v$ and $u$ components are somewhat less affected (18 versus 14~km~s$^{-1}$
and 25 versus 22~km~s$^{-1}$, respectively). Indeed, in all the cases the estimated dispersion
(for $n\ge 100$) is roughly equal to the sum in quadrature of the true dispersion and the
linear velocity error corresponding to the proper motion uncertainty. This is a useful
rule-of-thumb to estimate the expected bias of the PM under more general conditions.
A more unexpected result is that, for the smallest
samples ($n = 30$), the resulting dispersion matrix is unphysical (not positive definite)
in about 1\% of the cases. However, this does 
not happen with the larger samples.

The column to the right in Fig.~\ref{fig:dispersionvsuncertainty}
show the results of the ML method. We note that: 
\begin{enumerate}
\item Unlike the PM, the ML method is able to give unbiased dispersions even 
for very large $\sigma_{\mu}$, at least if the sample is not too small
($n\ge 100$). 
\item For the smallest samples ($n=30$) it is often necessary to use 
regularization. In most of these cases
it is sufficient to use $\alpha = 0.5$, which does not change the
axis ratio much. Only in very few cases was $\alpha > 3$ needed,
which gives a nearly isotropic dispersion.
\end{enumerate}
For the smallest sample size both PM and ML tend to underestimate
the dispersion, especially along the major axis ($u$ component) of the velocity 
ellipsoid.

%                                                                                      Two column figure
%________________________________________________________________

%________________________________________________________________

\subsection{Application to the Geneva--Copenhagen survey}

The Geneva--Copenhagen survey \citep{2004A&A...418..989N}
is the largest and most complete study of
a magnitude-limited, kinematically unbiased sample of F and G stars in the
solar
neighbourhood. It contains space velocities, Str{\"o}mgren
photometry, metallicities, rotation velocities and ages for more than
16\,000 stars. The catalogue gives complete $(u,v,w)$ space velocities
based on \emph{Hipparcos} data and radial velocities, from which the ``true''
vales of the mean velocity and velocity dispersion can be computed using
Eq.~(\ref{e00})

%________________________________________________________________

\subsubsection{Data used}

Not all stars in the Geneva--Copenhagen survey could be used in this
study.  Stars with large relative
parallax errors ($p/\sigma_{p}<10$), large
peculiar tangential velocities ($|\Delta\tau|>200$~km~s$^{-1}$), or
missing some of the necessary data were eliminated.
The remaining 7796 stars were divided into 10 bins of
approximately equal size, sorted by their Str{\"o}mgren color index
$b-y$.

%________________________________________________________________

\subsubsection{Results}

\begin{figure*}[!ht]
\begin{center}
\centerline{
\includegraphics[width= 0.65\columnwidth]{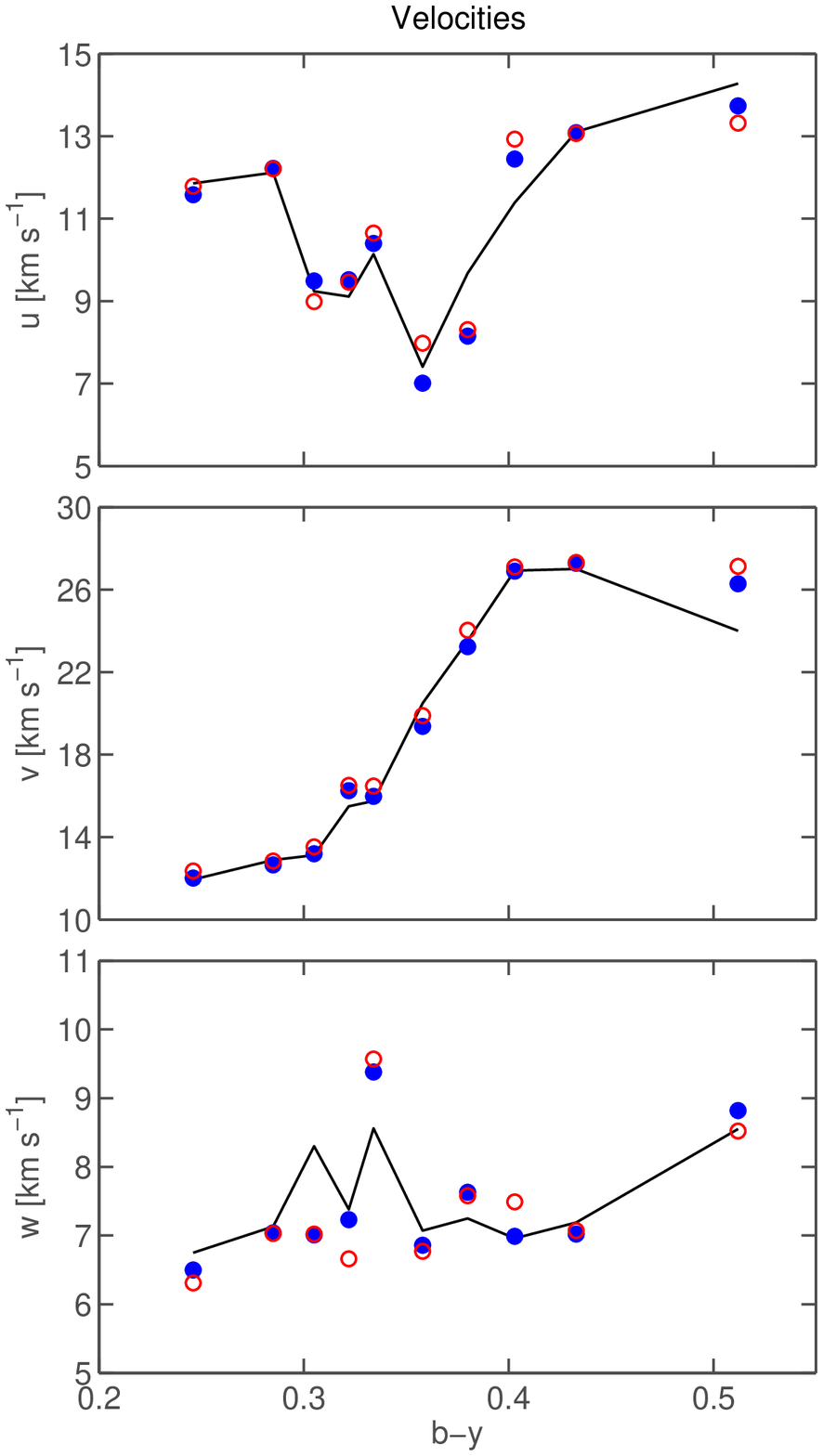}
%\hspace{0.01cm}
\includegraphics[width= 0.65\columnwidth]{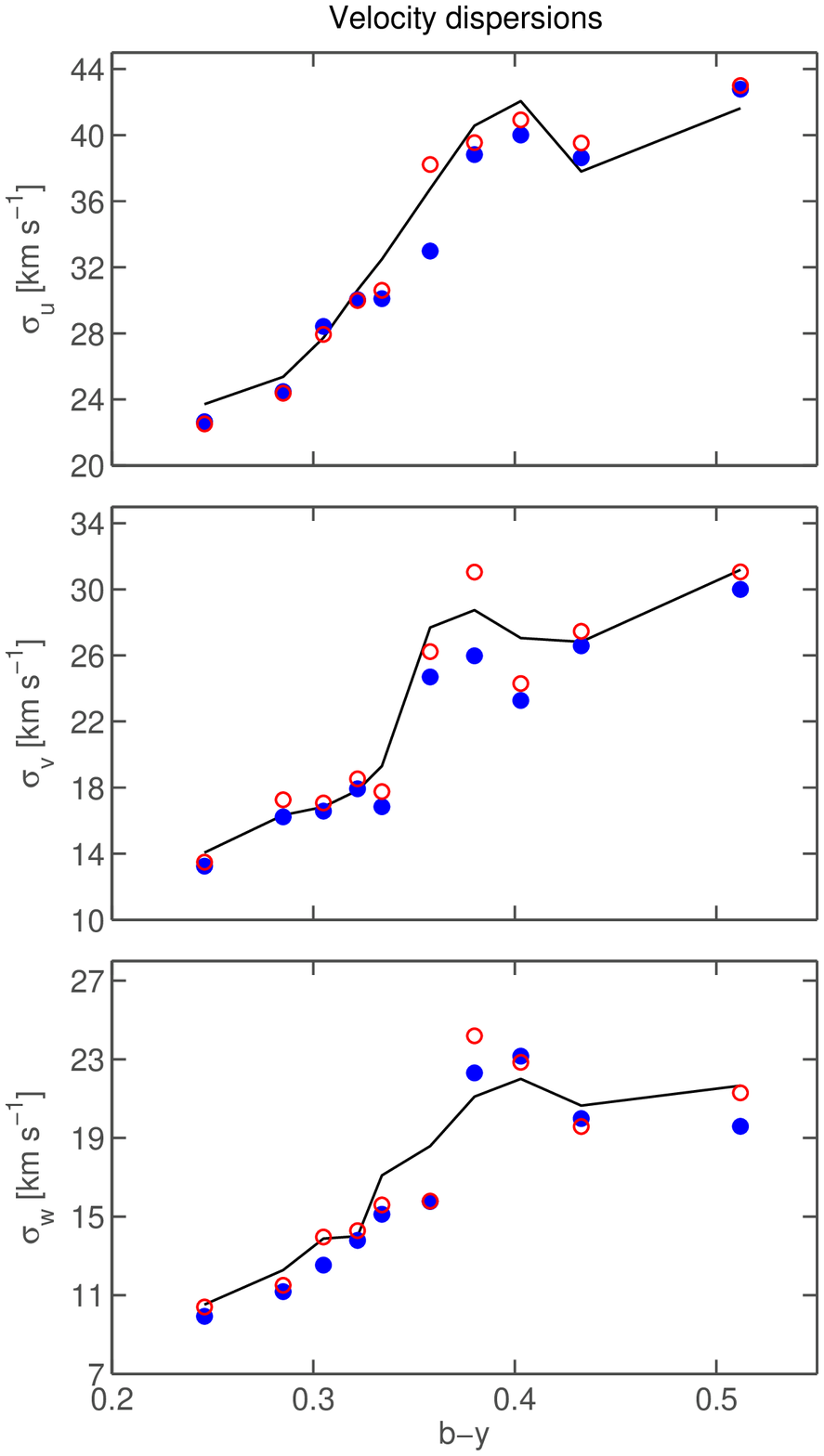}
%\hspace{0.01cm}
\includegraphics[width= 0.65\columnwidth]{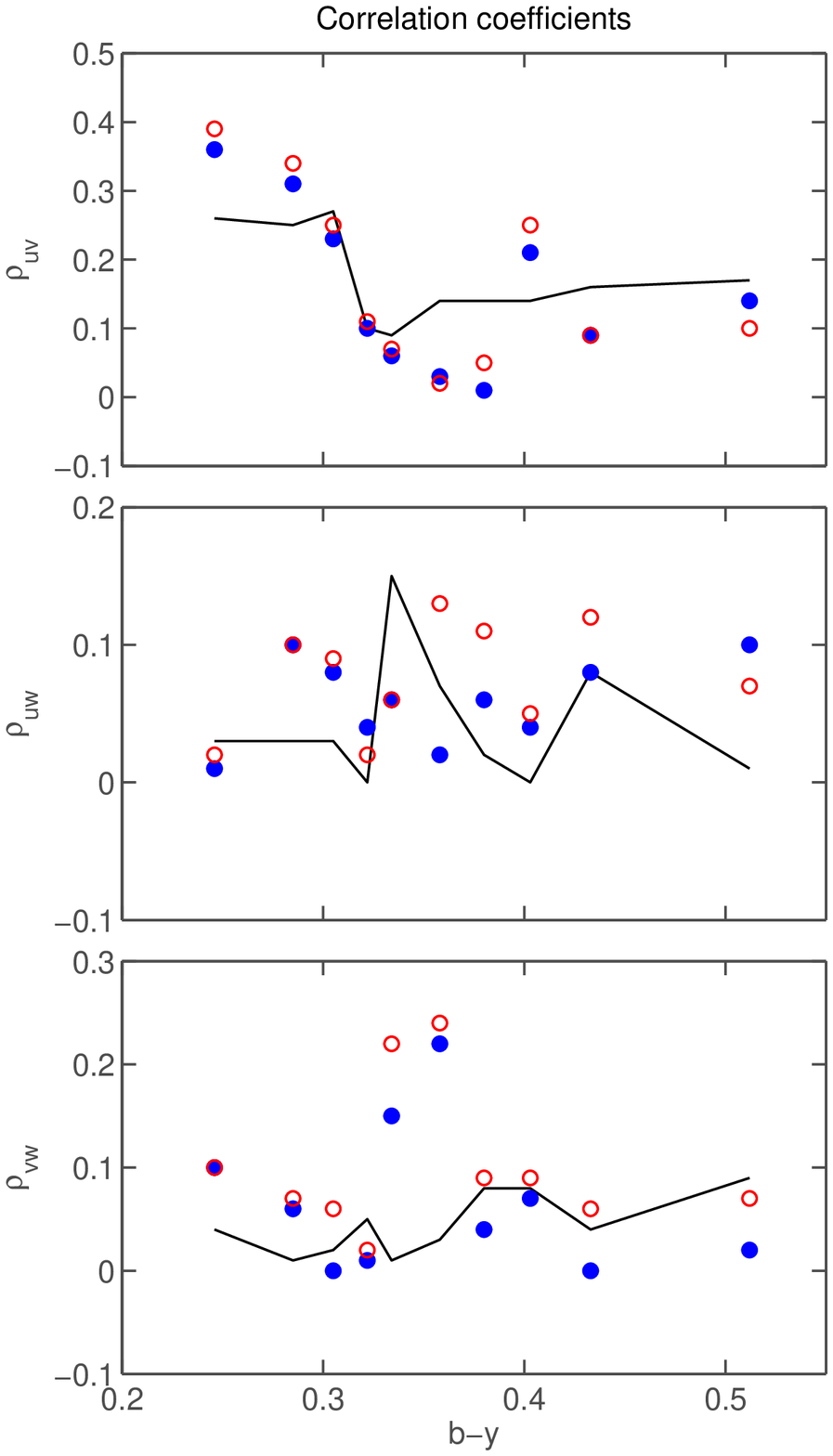}
}
\end{center}
\vspace{-.8cm}
\caption{\emph{Left.} Mean velocities $\overline{u},\overline{v},\overline{w}$ versus color index $b-y$ for stars in the
  Geneva-Copenhagen (GC) survey. \emph{Middle.} Velocity dispersions
  $\sigma_u, \sigma_v, \sigma_w$ versus color index $b-y$ for stars in the
 GC survey. \emph{Right.} Correlation coefficients $\rho_{ij} =
 D_{ij}(D_{i}D_{jj})^{-1/2}$ (where $i,j = u,v,w$) from the estimated
 dispersion matrix versus color
 index $b-y$ for stars in the GC survey. Values calculated directly from
  the GC data are shown as a solid line. The symbols represent the
  two different methods used: PM ($\circ$), ML  ($\bullet$).
}\label{fig:PMMLGC}
\end{figure*}

Table~\ref{GCDATA} and Fig.~\ref{fig:PMMLGC} give results of both estimation
methods (PM and ML) using tangential velocities only, as well as from the
full space velocities (GC). No error overlines are given here as the computation of the 
formal errors has not been implemented (cf.\ Sect.~\ref{sec:concl}).
  It is seen that PM and ML give very similar results for all the samples
  and for all nine kinematic parameters, and they both 
  agree well with the ``true'' values (GC). In the light of the experiments with 
  the synthetic samples this is not surprising since the 
observational errors in this case are generally small (few mas), resulting 
in almost negligible tangential velocity errors.
However, it is gratifying to note that the ML performs at least as well
as the PM in this case.

%________________________________________________________________

\section{Conclusions}
\label{sec:concl}

Galactic astrophysics is entering a new era in which large-scale
astrometric, photometric and spectroscopic surveys provide vast
quantities of data that need to be interpreted in terms of different
structural, kinematic and evolutionary models. In this context it is
essential to have at our disposal estimation techniques that can
deal with incomplete data (e.g., missing radial velocities) and that
are statistically efficient and reasonably unbiassed. It is also
important to take the observational uncertainties into account, for
although some of the surveys are much more precise than earlier ones
(e.g., \emph{Gaia} versus \emph{Hipparcos}), they are also applied to much more
distant stars, making the observational errors no less important.

%                                                                                  Two column table
%________________________________________________________________

\begin{table*}[t]
\caption{Results of both the projection method (PM)
  and our maximum likelihood method (ML) applied to samples from the Geneva--Copenhagen
survey \citep{2004A&A...418..989N}.}
\centering
\resizebox{.8\textwidth}{!}{
\begin{tabular}{cccrrrrrrrrr}
\toprule
$b-y$ & $n$ & Method & \multicolumn{1}{c}{$\overline{u}$} & \multicolumn{1}{c}{$\overline{v}$} & \multicolumn{1}{c}{$\overline{w}$} & \multicolumn{1}{c}{$\sigma_u$} & \multicolumn{1}{c}{$\sigma_v$} & \multicolumn{1}{c}{$\sigma_w$} & \multicolumn{1}{c}{$\rho_{uv}$} & \multicolumn{1}{c}{$\rho_{uw}$} & \multicolumn{1}{c}{$\rho_{vw}$} \\
\cmidrule{1-12}

\multirow{3}{*}{$0.205-0.272$} & \multirow{3}{*}{683} 
& GC & $-11.86$ & $-11.94$ & $-6.75$ & 23.72 & 14.06 & 10.52 &
0.26 & $-0.03$ & $-0.04$ \\ 

& & PM & $-11.79$ & $-12.36$ & $-6.31$ & 22.52 & 13.49 & 10.41 & 0.39 & 0.02 & $-0.10$ \\ 
& & ML & $-11.58$ & $-12.01$ & $-6.50$ & 22.65 & 13.24 & 9.93 &
0.36 & $-0.01$ & $-0.10 $\vspace{6pt}\\

\multirow{3}{*}{$0.272-0.296$} & \multirow{3}{*}{719} 
& GC & $-12.12$ & $-12.89$ & $-7.13$ & 25.37 & 16.33 & 12.28 & 0.25 & 0.03 & $-0.01$ \\ 
& & PM & $-12.21$ & $-12.84$ & $-7.03$ & 24.38 & 17.26 & 11.52 & 0.34 & 0.10 & 0.07 \\ 
& & ML & $-12.22$ & $-12.65$ & $-7.04$ & 24.47 & 16.23 & 11.19 &
0.31 & 0.10 & $-0.06$\vspace{6pt}\\ 

\multirow{3}{*}{$0.296-0.313$} & \multirow{3}{*}{704} 
& GC & $-9.24$ & $-13.12$ & $-8.30$ & 27.73 & 16.83 & 13.88 & 0.27 & $-0.03$ & $-0.02$ \\ 
& & PM & $-8.99$ & $-13.53$ & $-7.02$ & 27.93 & 17.07 & 13.96 &
0.25 & $-0.09$ & $-0.06$ \\ 
& & ML & $-9.49$ & $-13.19$ & $-7.01$ & 28.42 & 16.58 & 12.53 &
0.23 & $-0.08$ & 0.00\vspace{6pt}\\ 

\multirow{3}{*}{$0.313-0.331$} & \multirow{3}{*}{702} 
& GC & $-9.11$ & $-15.49$ & $-7.38$ & 30.62 & 17.83 & 14.00 & 0.10 & $-0.00$ & 0.05 \\
& & PM & $-9.46$ & $-16.51$ & $-6.66$ & 29.99 & 18.53 & 14.28 & 0.11 & 0.02 & $-0.02$ \\ 
& & ML & $-9.52$ & $-16.25$ & $-7.23$ & 30.04 & 17.93 & 13.79 &
0.10 & 0.04 & $-0.01$\vspace{6pt}\\ 

\multirow{3}{*}{$0.331-0.348$} & \multirow{3}{*}{768} 
& GC & $-10.14$ & $-15.75$ & $-8.56$ & 32.49 & 19.30 & 17.09 & 0.09 & $-0.15$ & 0.01 \\
& & PM & $-10.65$ & $-16.48$ & $-9.57$ & 30.60 & 17.75 & 15.60 & 0.07 & $-0.06$ & 0.22 \\
& & ML & $-10.40$ & $-15.98$ & $-9.38$ & 30.10 & 16.84 & 15.12 &
0.06 & $-0.06$ & 0.15\vspace{6pt}\\ 

\multirow{3}{*}{$0.348-0.370$} & \multirow{3}{*}{777} 
& GC & $-7.41$ & $-20.49$ & $-7.07$ & 36.75 & 27.70 & 18.59 & 0.14 & $-0.07$ & $-0.03$ \\ 
& & PM & $-7.98$ & $-19.89$ & $-6.77$ & 38.22 & 26.29 & 15.79 & 0.02 & $-0.13$ & 0.24 \\
& & ML & $-7.01$ & $-19.37$ & $-6.86$ & 32.98 & 24.69 & 15.77 &
$-0.03$ & 0.02 & 0.22\vspace{6pt}\\ 

\multirow{3}{*}{$0.370-0.391$} & \multirow{3}{*}{820} 
& GC & $-9.68$ & $-23.52$ & $-7.25$ & 40.57 & 28.75 & 21.10 & 0.14 & $-0.02$ & $-0.08$ \\ 
& & PM & $-8.31$ & $-24.03$ & $-7.58$ & 39.55 & 31.05 & 24.19 &
$-0.05$ & $-0.11$ & $-0.09$ \\
& & ML & $-8.15$ & $-23.24$ & $-7.63$ & 38.83 & 25.98 & 22.30 &
$-0.01$ & $-0.06$ & $-0.04$\vspace{6pt}\\

\multirow{3}{*}{$0.391-0.416$} & \multirow{3}{*}{835} 
& GC & $-11.39$ & $-26.93$ & $-6.96$ & 42.07 & 27.06 & 22.00 & 0.14 & 0.00 & $-0.08$ \\ 
& & PM & $-12.93$ & $-27.10$ & $-7.49$ & 40.93 & 24.30 & 22.84 & 0.25 & 0.05 & $-0.09$ \\ 
& & ML & $-12.45$ & $-26.90$ & $-6.99$ & 40.01 & 23.28 & 23.16 &
0.21 & 0.04 & $-0.07$\vspace{6pt}\\ 

\multirow{3}{*}{$0.416-0.454$} & \multirow{3}{*}{865} 
& GC & $-13.11$ & $-27.00$ & $-7.19$ & 37.81 & 26.82 & 20.64 & 0.16 & $-0.08$ & 0.04 \\
& & PM & $-13.07$ & $-27.32$ & $-7.07$ & 39.52 & 27.46 & 19.58 & 0.09 & $-0.12$ & 0.06 \\ 
& & ML & $-13.09$ & $-27.28$ & $-7.02$ & 38.63 & 26.58 & 19.99 &
0.09 & $-0.08$ & 0.00\vspace{6pt}\\ 

\multirow{3}{*}{$0.454-0.981$} & \multirow{3}{*}{923} 
& GC & $-14.28$ & $-24.00$ & $-8.55$ & 41.63 & 31.18 & 21.66 & 0.17 & $-0.01$ & 0.09 \\ 
& & PM & $-13.32$ & $-27.13$ & $-8.52$ & 43.00 & 31.06 & 21.29 & 0.10 & $-0.07$ & 0.07 \\ 
& & ML & $-13.74$ & $-26.29$ & $-8.82$ & 42.78 & 30.00 & 19.59 &
0.14 & $-0.10$ & $-0.02$ \\

\bottomrule
 \end{tabular}}
\tablefoot{
The stars are divided into color bins according to
the limits in $b-y$ in the first column. $n$ is the number of stars in the bin. The third
column shows the estimation method used. The subsequent columns give
the mean velocities $\overline{u}$, $\overline{v}$, $\overline{w}$, the velocity dispersions $\sigma_u$, $\sigma_v$, $\sigma_w$ 
(in km~s$^{-1}$) and the correlation
coefficients $\rho_{ij}=D_{ij}(D_{ii}D_{jj})^{-1/2}$ (where $i,j=u,v,w$) from the estimated dispersion matrix. GC means using the full space velocities,
including the radial velocities as given by
\citet{2004A&A...418..989N}.
}
 \label{GCDATA}
\end{table*}

For many stars brighter than 17th magnitude \emph{Gaia} will provide 
the complete space velocity vector through a combination of its astrometric
and spectroscopic measurements. However, the vast majority of stars
observed by \emph{Gaia} are fainter than 17th magnitude and will have no radial
velocities. This is also the magnitude range where the
astrometric errors in parallax and proper motion start to become 
problematic for studies of the galactic kinematics beyond the solar
neighbourhood. For example, solar-type stars at 5~kpc distance
will be observed at apparent magnitude 18 to 20, depending on the
extinction, resulting in relative parallax errors of at least 50\% and 
transverse velocity errors of several km~s$^{-1}$. The very large number
of such stars will permit a detailed mapping of their distribution
functions provided that the statistical biases can be mastered.

In this paper we have considered a seemingly very simple problem,
namely to estimate the nine parameters describing the velocity
distribution of stars under the Schwarzschild approximation, based
exclusively on astrometric data and taking into account the
observational uncertainities in the parallaxes and proper motions. The
rigorous application of the maximum likelihood method to this problem
turns out to be surprisingly complex, and we have devised an
approximate numerical method to solve it. We have tested the method on
both synthetic and real samples of stellar data, and found that it
performs slightly better than the simple projection method \citep{DB98}, 
especially when the observational errors are important
and for small samples. For very small samples (less than about 30
stars) the projection method sometimes gives unphysical results. This
is avoided with our method, which however may require regularization
for such small samples, leading to a more isotropic dispersion tensor.

%                                                                                  Two column figure
%________________________________________________________________

Several authors have applied maximum likelihood or Bayesian methods to
estimate stellar kinematics using formulations that differ in various
aspects from our approach. For example, \citet{1998AJ....115.2384D}
describes how general three-dimensional velocity distributions can be
derived from tangential velocities, using the so-called maximum
penalized likelihood estimate, i.e., a maximum-likelihood estimate
constrained to give a smooth velocity distribution. While this method
is very general, it does not take into account the observational
uncertainties. \citet{2005ApJ...629..268H} and
\citet{2009ApJ...700.1794B} derive the parameters for a mixture of
Gaussian distributions from tangential velocity data. While they do
include the observational uncertainties through a linear propagation
to the tangential velocities, their formalism is quite different from
ours in the treatment of the parallax errors. It could be interesting
to compare the two methods as they both depend on (different) 
approximations for small relative parallax errors.

The present method could be extended and generalized in various
ways. One obvious extension is to allow that some stars have a
measured radial velocity instead of, or in addition to the astrometric
data. Another is to consider a more complex velocity distribution,
e.g., a superposition of several Gaussian components. It is also
desirable to apply a more efficient numerical solution technique than
the downhill simplex method which was chosen for the
present experiments for its ease of implementation. Alternative 
methods such as a quasi-Newton with finite differences \citep{book:nr3} 
would almost certainly be much more efficient.
We note that the computing time scales linearly with the number of
stars (for a fixed number of kinematical parameters), which makes it
feasible to apply the ML method also to much larger samples than in 
our experiments. Finally, the formal errors (or more generally confidence 
regions) of the estimated quantities need to be computed, which could 
be done with moderate effort by mapping the log-likelihood function
around its maximum.

%________________________________________________________________

\begin{acknowledgements} We wish to thank the anonymous referee for
  several useful suggestions which helped to improve the
  manuscript.\end{acknowledgements}

%________________________________________________________________

\bibliography{refs}
\bibliographystyle{aa}

\end{document}